# On the accumulation of deleterious mutations during range expansions

Stephan Peischl[1,2,3], Isabelle Dupanloup[1,3], Mark Kirkpatrick[2], and Laurent Excoffier[1,3]

[1] Institute of Ecology and Evolution, University of Berne, 3012 Berne, Switzerland

[2] Section of Integrative Biology, University of Texas, Austin Texas 78712, USA

[3] Swiss Institute of Bioinformatics, 1015 Lausanne, Switzerland

Corresponding authors:

Stephan Peischl
Institute of Ecology and Evolution
University of Bern
Baltzerstrasse 6
CH-3012 Bern
Switzerland
Phone: +41 31 631 30 36
Fax: +41 31 631 48 88
Email: stephan.peischl@iee.unibe.ch

Laurent Excoffier
Institute of Ecology and Evolution
University of Bern
Baltzerstrasse 6
CH-3012 Bern
Switzerland
Tel: +41 31 631 30 31
Fax: +41 31 631 48 88
Email: laurent.excoffier@iee.unibe.ch

# Abstract

We investigate the effect of spatial range expansions on the evolution of fitness when beneficial and deleterious mutations co-segregate. We perform individual-based simulations of a uniform linear habitat and complement them with analytical approximations for the evolution of mean fitness at the edge of the expansion. We find that deleterious mutations accumulate steadily on the wave front during range expansions, thus creating an *expansion load*. Reduced fitness due to the expansion load is not restricted to the wave front but occurs over a large proportion of newly colonized habitats. The expansion load can persist and represent a major fraction of the total mutation load thousands of generations after the expansion. Our results extend qualitatively and quantitatively to two-dimensional expansions. The phenomenon of expansion load may explain growing evidence that populations that have recently expanded, including humans, show an excess of deleterious mutations. To test the predictions of our model, we analyze patterns of neutral and non-neutral genetic diversity in humans and find an excellent fit between theory and data.

# Introduction

The range of most species is fluctuating, a phenomenon that now occurs at an increasing rate owing to rapid climatic changes [e.g. 1,2,3]. Many successful species have increased their range just after speciation [4], and climatic changes can trigger expansions, shifts and contractions for temperature sensitive species [5,6,7]. Understanding the theoretical and practical implications of dynamic range margins for the ecology, biology, and genetics of species has become an important topic in evolutionary biology.

Range expansions can promote the spread of new and standing mutations that happen to be on the wave front of an expansion [8], a phenomenon called *gene surfing* [9]. By surfing on an expanding wave front, a neutral mutation can quickly fix and spread over large regions, mimicking the effect of adaptation to new environments. Surfing is due to enhanced genetic drift on the expansion wave front, where population density is low and growth rate is high. It can lead to clines of heterozygosity from the source to the edge of the expansion [10,11,12,13], which has been invoked to explain the decline of genetic diversity with distance from Africa in humans [14,15,16,17]. Surfing is not restricted to neutral mutations: both beneficial and deleterious mutations can also surf [18,19,20].

Most previous theoretical and empirical studies have focused on the consequences of range expansions at single loci, and little is known about situations when multiple loci segregate. Simulations of a haploid two-locus model with epistasis showed that range-shifts can substantially increase the chance of crossing a fitness valley [21]. It remains unclear how range expansions affect adaptation under a flux of beneficial and deleterious mutations that appear simultaneously throughout the genome.

In this paper, we investigate the effect of range expansions on the evolution of fitness in different parts of a species range and on the mutation load. Our model includes both deleterious and advantageous mutations. We use both individual-based simulations and analytical approaches to describe this process. We then examine patterns of genetic variation from human populations, and find that they are qualitatively consistent with our predictions.

## Results

### Fitness decreases during range expansions

We first performed individual-based simulations of a range expansion along an unbounded linear habitat. Individuals are diploid and, for sake of simplicity, their genome is structured into *n* non-recombining segments that segregate freely. Unless stated otherwise, each segment is subject to both advantageous and deleterious mutations of fixed effect sizes *s* and -*s*, respectively. Fitness effects are multiplicative such that each mutant allele carried by an individual increases (decreases) its fitness by a factor $1+s$ ($1-s$), that is there is no dominance or epistasis. We assume in the following that 90% of all non-neutral mutations are deleterious, which seems conservative [22], but our results can be generalized for arbitrary distributions of fitness effects (see Figures S1 and S2). Figure 1A shows the evolution of the mean fitness during a linear range expansion. We see a very strong difference in the rate of adaptation between core and peripheral populations. Deleterious mutations accumulate faster than beneficial mutations on the wave front, leading to a steady decrease in mean fitness during the expansion. In contrast, recurrent selective sweeps increase the fitness of core populations. This difference between the periphery and core generates a cline in fitness whose steepness increases over time. By tracking the spatial origin of mutations, we find that a large fraction of the total load is due to deleterious mutations that originated on the wave front (Figures 1B and S3). We call this fraction of the mutation load the *expansion load*.

Expansion loads is only prevented when deleterious mutations have a large effect, but not necessarily by increasing migration rates when mutations have a small effect (Figure S4). Mean fitness at the wave front decreases fastest when migration is small and selection is intermediate (Figure S4). As expected, the rate at which mean fitness declines for a given combination of parameters decreases with increasing local carrying capacities, which make selection more operational (cf. Figures S4 C and D). While increasing the rate of migration has almost no influence on the build-up of expansion load if selection is weak, it decreases the rate at which fitness declines if selection is intermediate or strong. In summary, expansion load is expected to occur for selection coefficients of up to a few percent and *local* carrying capacities of several hundred to thousands of individuals per deme (see Figure S4 and S5).

**Expansion load affects a large part of the species range**

While the expanding wave front represents at any time only a tiny part of a species' entire range, most populations of an expanding species descend from ancestors who reproduced at the front [23]. Consequently, expansion load can represent a substantial part of the total load for most of the demes of an expanding species. For instance, in the case presented in Figure 1B, more than 50% of the total load is due to mutations that originated at the wave front in territories colonized 300 generations after the onset of the expansion.

We can better appreciate this phenomenon by looking separately at the distribution of deleterious and beneficial mutations in space. Figure 2 shows that the number of deleterious mutations per deme increases almost linearly along the expansion axis, whereas the number of beneficial mutations decreases with distance from the origin of the expansion. This pattern is quite stable over time as similar gradients in deleterious and beneficial mutations are observed 1000 and 2000 generations after the onset of the expansion (Figure 2). The slow purging of deleterious mutations suggests that they are often at high frequency within demes.

Although these results pertain to expansions in an unbounded range, a very similar pattern occurs on a finite habitat that is 200 demes wide colonized in about 800 hundred generations (Figure S6). The fitness clines visible 1000 and 2000 generations after the onset of the expansion are very similar to those in an unbouded habitat, and the clines even persist 5000 generations after the onset of the expansion (Figure S6). The main difference with an unbounded expansion is in the presence of an edge effect on the righthand side of the habitat because the influx of beneficial mutations into the rightmost demes is reduced.

## Analytical approximations

In order to better understand how expansion load builds up, we derived an analytical approximation for the expected change of mean fitness at the wave front in a simplified expansion-model (see *Material and methods*). Using standard diffusion approximations (e.g. [24]), we find that the probability of fixation of a new mutation at the wave front is

$$p \approx \frac{\exp(2\,s_e) - 1}{\exp(4 s_e K) - 1},  \qquad [1]$$

where $s_e = s\,m\log(2/m)/(2r)$ is an effective selection coefficient. Note that Eq. [1] is equivalent to the probability of fixation of a mutation with effect $s_e$ in a single panmictic population of constant size $K$ [24] (see Figure S7). The fact that $|s_e| < |s|$ (Figure S8) shows that selection is always less efficient in expanding than in stationary populations, which is because the expansion process increases the strength of drift at the wave front. Deleterious mutations will thus become more readily established at the wave front and beneficial mutations will sweep less often than in the core, contributing to the building up a mutation load at the front. It also follows from Eq. [1] that the efficiency of selection at the wave front decreases with increasing growth rate, and increases with increasing migration rate (Figure S8). Higher growth rates accelerate the expansion and hence the strength of drift at the wave front [10]. Increasing migration rates decrease the severity of founder effects and of drift during the colonization of new demes.

Eq. [1] can also be used to infer the change in mean fitness at the wave front, by assuming that it is due to the serial fixation of independent mutations. Unless selection is strong (i.e. $K\,s \gg 1$), our analytical approximation is in excellent agreement with individual-based simulations done under a more complex expansion model in one dimension (relative error of less than 5%, see Figures S1, S2, and S9 - S12), but also in 2D expansions (see Figure 3 below). Overall, we find that our analytical approximation is conservative in the sense that it tends to underestimate the parameter region in which expansion load occurs (cf. Figure S4 A and C, and B and D).

## Expansion load in two-dimensional expansions

We find that deleterious mutations accumulate at the same average rate during linear or radial expansions in 2D habitats than in 1D expansions. Indeed, as shown on Figure 3, the evolution of the mean fitness at the wave front in linear or radial 2D expansions is remarkably similar to the 1D case. This suggests that 1D analytical theory for the evolution of mean fitness at the wave front can be applied to 2D expansions. The variance in fitness at a given distance from the source of the expansion increases, however, with the width of the front (Figure 4). Mean fitness can even increase in some parts of the wave front during early stages of the expansion before it eventually decreases over time (Figure 4). We also observe that in 2D habitats populations recover more quickly from expansion load (i.e., deleterious mutations are purged more quickly, compare panels A and C in Figure 2) than after a 1D expansion. This difference is likely due to several causes. First, different deleterious mutations accumulate in different parts of the 2D wave front and single deleterious mutations rarely fix on the whole wave front, unlike in 1D expansions. Second, stationary 2D structured populations have a larger effective size making selection more efficient after the expansion [25]. Indeed, because the total population size is larger in the 2D model, the influx of beneficial mutations and hence the total number of established beneficial mutations increases relative to 1D expansions (see panels B and D in Figure 2). Nevertheless, despite a faster recovery, expansion load can still be visible for thousands of generations after the expansion has stopped (see Figures S13 B and S14).

## Evidence for expansion load in humans

### Non-African populations show an excess of deleterious mutations

There is mounting evidence that human populations that expanded out of Africa carry an excess of rare deleterious mutations [e.g., 26,27,28,29]. We ask here if this pattern might result from an expansion load as predicted by our model. To address this question, we first analyzed the autosomal exomic diversity in 17 Africans and in 25 non-Africans sequenced at high coverage (>50X, [30], see *Material and Methods* for details). In agreement with a previous analysis of a smaller representation of functional human diversity [27], we find that the proportion of deleterious mutations is significantly higher in non-Africans (Table 1). This excess is particularly strong when focusing on private derived alleles, which are likely to have arisen recently. More than 27% of private alleles in non-African populations are predicted to have deleterious effects, as compared to only 21% in Africans (*p* < 0.001, based on a permutation test). These results are in agreement with our model, which suggests that many of the deleterious mutations in Europeans arose during the range expansions out-of-Africa. An alternative explanation for the observed pattern would be an excess of rare genetic variants, caused by a recent explosive human population growth [31,32,33]. However, the observed excess of deleterious mutations observed in large samples was mainly due to very rare or even private deleterious mutations, whereas we find that the excess of deleterious mutations in non-Africans is not restricted to rare variants (Table 1). We therefore suggest that (some of) these deleterious mutations may be the signature of expansion load outside Africa.

**Diversity at non-neutral sites decreases linearly with distance from Africa**

Human neutral genetic diversity has been shown to decrease with increasing distance from Africa [14,16,34], which has been interpreted as being due to recurrent founder effect during range expansion [13,15,17]. We compare here the spatial pattern of genetic variation at neutral and selected sites using a large set of SNPs [34] genotyped in 53 worldwide populations [35]. We obtained the evolutionary conservation index (GERP score, [36]) for >655,000 SNPs (see *Material and Methods*). Sites with a GERP score between -2 and 2 were considered as neutral, whereas those with a GERP score larger than 4 we considered as under strong purifying selection. For each population, we then calculated the average expected heterozygosity $H_n$ for neutral SNPs, and $H_s$ for sites under purifying selection.

Figure 5 shows $H_n$ and $H_s$ plotted against distance from Ethiopia, a putative origin of the human expansion out of Africa [14,16]. As expected, we find that $H_n$ decreases with distance from Ethiopia (Figure 5A), and that $H_s$ is smaller than $H_n$ in all populations (Figure 5C), the latter observation being consistent with purifying selection decreasing the frequencies of deleterious alleles [37]. We find that $H_s$ also decreases with distance from Ethiopia (Figure 5B), and that the slopes of the clines in $H_s$ and $H_n$ are similar outside Africa (-4.5 x 10$^{-6}$/km and -4.7 x 10$^{-6}$/km, respectively). Analysis of covariance shows that the difference between the slopes is not significant ($p > 0.7$). This suggests that similar evolutionary processes have affected neutral and selected variants during recent human range expansions.

We then used the reduction in heterozygosity at conserved sites $RH = (H_n - H_s)/H_n$ as a proxy for the efficiency of selection (a stronger reduction indicates more efficient selection against deleterious mutations [38]), to check that selection is globally more efficient in African populations. *RH* is found significantly larger (t-test, $p = 10^{-7}$) in Africa (9.7 % +- 0.3 SD) than outside of Africa (4.3% +- 0.5 SD, Figure 5C), in keeping with the hypothesis of a reduced selection outside Africa. Interestingly, in non-African populations *RH* does not depend on their distance from Africa and is similar in all populations (Figure 5C). This observation is not compatible with the existence of a cline in effective population size that would condition the (equilibrium) level of neutral and selected heterozygosity. In this case one would indeed expect to see a cline in *RH*. The uniform reduction of heterozygosity outside of Africa rather suggests that sites under purifying selection in Africa have been evolving neutrally during the expansion out of Africa and that their allele frequencies have been mainly shaped by the range expansion. The smaller average heterozygosity of highly conserved sites simply reflects their smaller initial minor allele frequencies in the populations leaving Africa.

**Comparison with simulations**

To check if a simple model of range expansion can reproduce the observed heterozygosity patterns, we simulated the evolution of neutral and deleterious mutations during range expansions and recorded their average expected heterozygosities (see *Material and Methods* for details on the simulations). By varying migration rates and selection coefficients we identified parameter values that fit the observed clines in heterozygosity very well (*m* = 0.1, *s* = 0.002; see Figure 5). Although the fit between simulated and observed data was only based on the cline of neutral heterozygosity outside Africa and the reduction of heterozygosity in Africa, simulations reproduce two additional features seen in the data. First, we obtain an excellent agreement for the cline of heterozygosity at conserved sites outside Africa (Figure 5B). Second, we can reproduce the smaller and uniform reduction of heterozygosity outside Africa for conserved loci (Figure 5C) (simulated mean *RH* = 10.7% +- 0.1 SD in Africa vs. 4.7% +- 0.8 SD out of Africa, *p* = $10^{-17}$, t-test). Note that the overshoot of simulated heterozygosity in African populations (cf. Figure 5 A and B) is presumably due to an unaccounted source of ascertainment bias in our data (see *Material and Methods* for details on our correction for ascertainment bias).

## Discussion

Evolution during range expansions is often viewed as a process in which species evolve adaptively as they encounter new environments [39,40]. While adaptation certainly plays a role, there is evidence that deleterious mutations are also established [27,41,42]. We show here that genetic drift on the front of range expansions can lead to a steady and long-lasting accumulation of deleterious mutations over most of a species range, a phenomenon we call expansion load. This load develops under quite general conditions, such as large local carrying capacities (>1000 individual, Figure S5), large selection coefficients (up to several percent, Figure S4), large migration rates (>25%, Figure S4), long distance dispersal (Figure S9), or alternative distributions of fitness effects (Figure S1). The most important assumption of our model is that deleterious mutations occur frequently relative to beneficial mutations. For instance, we can use Eq. [5] to find the minimum proportion of selected mutations to be deleterious to still have expansion load. For the parameter values used in Figure 1, we estimate that if at least 57% of selected mutations are deleterious, fitness will decrease along the expansion axis (see Figures S2 and S15).

Stronger migration usually decreases the rate at which mean fitness declines because bottlenecks during colonization events are less severe and peripheral populations are less isolated from the core. Note, however, that even very large migration rates between neighboring demes (i.e. *m* > 0.2) cannot completely prevent an expansion load from developing unless the fitness effects of mutation are strong (*s* > 0.02) and the deme carrying capacities are large (*K* > 200; see Fig. S4). This implies that unless species have large local carrying capacities (as with some invertebrates and microbes), they are very likely to be affected by expansion load and their fitness will decrease during the expansion. This decline is fastest when mutations have intermediate fitness effects (*s* between 1% and 5%, depending on parameter values, see Figure S4). Although the expansion load is progressively eliminated by selection in the range core, this process can be quite slow. Consequently, expansion load can linger as a major component of the mutational total load for thousands of generations (see Figures 1B, 2, S3, and S6). Note that these predictions also hold for linear or radial 2D expansions (see Figures 2, S13 B, and S14), despite the overall larger variance in fitness seen in 2D expansions (Figure 4).

## Evidence for expansion load

Bottlenecks and population expansions decrease the ability of selection to purge slightly deleterious mutations [27,29,31]. A bottleneck has been proposed as the explanation for the observed excess of non-synonymous or deleterious mutations in non-African populations [27]. Our simulations of a single bottleneck confirm that it can increase the number of deleterious mutations (Figure S16). However, this bottleneck needs to be unrealistically severe, with a very small population size and a very long duration to show levels of mutation load that are easily obtained after a relatively short range expansion (compare Figures 1B and S3 to Figure S16). This is in keeping with the results of Lohmueller et al. [27] who showed that a bottleneck lasting more than 7,500 generations (>150 Ky) is necessary to produce the excess of deleterious mutations observed in non-Africans. However, a single bottleneck cannot account for the cline in heterozygosity observed in humans (see e.g. [14,16,34], and Figure 5). In contrast, our model of expansion load can explain both the clines in heterozygosity at neutral and conserved sites outside Africa and the observed difference in selection efficiency between African and non-African populations.

It would be interesting to more explicitly test our model predictions by looking for a correlation between the extent of mutation load and distance from Africa, which would require high quality genomic data from a collection of well-chosen populations. It would also be interesting to estimate the (selection) parameters that best explain these data, such as to extrapolate the average fitness of the populations along the expansion axis. This parameter estimation process would be computationally very demanding if it were to be based on our current forward simulations integrated for instance in an Approximate Bayesian Computation framework (ABC, [43], and see [44] for a recent attempt at estimating selection coefficients with ABC). We note however, that the genomic study of more recent expansions in well-defined populations showing evidence for an increased rate of rare diseases (e.g., in Finland [45] and Quebec [46]) could also potentially reveal signals of expansion load.

Fewer genomic resources are available in other species than humans, which makes it harder to detect expansion load, but the solitary bee *Lasioglossum leucozonium* that recently expanded to North America might be a potential example. Invasive populations of this species indeed carry mutations that reduce population growth [47]. Another potential case comes from the flowering plant *Mercurialis annua*. Populations that recently invaded the Iberic peninsula from North Africa interestingly show no evidence of inbreeding depression, apparently because mildly deleterious mutations were fixed during the range expansion [48]. A third suggestive observation is that populations of several invasive species sometimes suddenly collapse without clear explanations (see, e.g., [49], for a review of 17 such cases). Our results suggest that those extinctions could have a genetic basis, a hypothesis that could be tested.

**What might prevent expansion load**

Many species have successfully colonized new areas, suggesting either that it is possible to escape the negative consequences of expansion load, that the reduction in fitness has not been severe enough to prevent and stop the expansion, or that expansion load has been compensated by beneficial mutations. Several factors not taken into account in our analyses might mitigate the load. Our model assumes that mutations affect fitness multiplicatively. Negative epistatic interactions among deleterious mutations could increase the purging of deleterious mutations and so mitigate the expansion load [50]. Similarly, non-random mating with respect to fitness has been shown to increase the efficiency of selection against deleterious mutations [51]. Allee effects (lower growth rate at low densities) have been shown to preserve genetic variation at the front of expanding populations [52] and thus increase effective population size on the front and lower the severity of expansion load.

Our results show that fitness variance at the wave front increases with the width of the wave front in 2D expansions (see Figure 4). Despite this spatial genetic variation at the wave front, it is remarkable that mean fitness at the wave front decreases at essentially the same rate as in 1D expansions (see Figure 3). This might be due to the fact that in 2D expansions, heterozygosity is essentially nil on the wave front (see Video 1 in the SI), even though it quickly recovers by gene flow from neighboring populations. The main difference between 1D and 2D expansions is that populations in the wake of the expansion recover their fitness more quickly in 2D than in 1D habitats (see Figure 2). Note that even though we restricted our study to uniform environments, we would posit that the observed variance in fitness on 2D expanding fronts can be useful for invading populations that need to adapt to new environmental conditions.

## Conclusions

The accumulation of deleterious mutations during range expansions has been largely unappreciated, perhaps because the main focus of most studies has been on adaptive processes. We show here that many species whose ranges have recently expanded are expected to suffer an increased mutation load. This excess of deleterious mutations is not restricted to the wave front, but affects all the newly colonized species range and can persist for thousands of generations after the end of the expansion. We find that an expansion load can explain a growing body of data from several species, including humans. However, since many species are successful invaders, our results do not imply that expanding species are doomed, but would suggest that mechanisms preventing expansion load might have been overlooked and deserve further studies.

## Material and Methods

### Simulation models

We model a range expansion as the successive colonization of empty demes located on a one- or two-dimensional lattice. Generations are discrete and non-overlapping. Migration is homogeneous and isotropic, except that the boundary is reflecting, i.e., individuals cannot migrate out of the habitat. Adults migrate between adjacent demes with probability *m*, and mating within each deme is random. Demes grow logistically such that the expected number of juveniles in the next generation is given by $N^* = \exp(r)N / [1 + (\exp(r) - 1)N / K]$, where $r$ is the maximal growth rate, and *K* the carrying capacity [53]. The actual number of offspring, *N'*, is then drawn from a Poisson distribution with mean *N\**. Mating pairs are formed by randomly drawing individuals (with replacement) according to their fitness. Each mating pair produces a single offspring and the process is repeated *N'* times, leading to an approximately Poisson-distributed number of offspring per individual.

Individuals are monoecious and diploid. Each gamete carries *k* new deleterious (beneficial) mutations per generation, where *k* is drawn from a Poisson distribution with mean $u_d$ ($u_b$). Ignoring neutral mutations, we denote the genome wide mutation rate $u = u_d + u_b$. Mutations are randomly distributed over *n* independently segregating regions. Within these regions, sites are completely linked and each new mutation falls on a unique site (infinite-site model). We denote by $\varphi_d = u_d / u$ the probability that a new mutation is deleterious and by $\varphi_b = 1 - \varphi_d$ the probability that it is beneficial.

Fitness effects are multiplicative, such that the fitness of an individual is given by $w = \prod_i (1 + s_i)$, where $s_i$ is the selection coefficient associated to the *i*-th mutation at any locus of the focus individual, i.e., there is no dominance or epistasis.

Before the onset of the expansion, the five leftmost demes (or five leftmost columns in 2D linear expansions) are at carrying capacity and all other demes are empty. Unless specified otherwise, the expansion starts after a burn-in phase of 10 *K* generations during which individuals migrate between already colonized demes but not into new territories. Mean fitness at the edge of the colonized area is then normalized to 1 at the onset of the expansion.

**Simulation of human range expansion**

To simulate the evolution of heterozygosity in human populations, we keep track of allele frequencies at a set of 500 selected and 500 neutral loci. All loci are assumed diallelic and unlinked. At selected loci, the ancestral allele is assumed selectively neutral and mutants reduces an individual's fitness by a factor 1-*s* only if it is present in homozygous state, that is, deleterious mutations are completely recessive. Because we are modeling mutations at single nucleotides, we assume the frequency of back mutation to be sufficiently rare that it can be neglected, and that each mutation occurs at a unique locus.

We modeled human range expansion to occur on an array of 10x100 demes, with an ancestral population restricted to the first 10x5 demes on the left of the habitat. After reaching migration-selection-drift equilibrium, the population are allowed to expand into the empty territory, which is separated from the ancestral population by a spatial bottleneck (to mimic the bottleneck out of Africa, see Figure S17 for an illustration of the model). After 2,000 generations (corresponding to about 50 Ky since the exit out of Africa [54,55]), we computed the average expected heterozygosity for all populations. To compare the simulation results with the data, the spacing of demes was chosen such that the slope of the linear regression of neutral expected heterozygosities of non-African populations against distance from Ethiopia matched the regression slope calculated from the observed data. Since computational limitations of individual-based simulations prohibit a complete exploration of the parameter space for this model, we focused on a set of reasonable demographic and mutations parameters ($K$ = 100 diploid individuals per deme, mutation rate of $u = 10^{-4}$ per locus per generation), and the migration rate and selection coefficient was adjusted to lead to a good fit with the observations.

**Correction for ascertainment bias of HGDP SNPs**

Some ascertainment occurred when selecting SNPs to be put on the Illumina Hap650Y Genotyping BeadChip used to genotype the HGDP samples SNPs [34], which are a subset of SNPs defined in HapMap samples potentially enriched for frequent tag SNPs in European samples, even though the exact ascertainment scheme is not precisely defined (see www.illuminakk.co.jp/pdf/HUMANHAP650Y_DataSheet.pdf). To mimic an ascertainment bias towards an excess of frequent SNPs in European samples, we sampled 20 individuals from the deme with coordinates (5,11), that is, in the expanding population, located 5 demes away from the migration barrier, and only loci that were polymorphic with a minor allele frequency larger than 5% in this small discovery sample were then used for the calculation of the average expected heterozygosity for each population. The average expected heterozygosity was then simply calculated as

$H = 1/n \sum_{i=1}^{n} 2p_i(1-p_i)$, where $i$ runs over the $n$ loci selected after the ascertainment bias correction, and $p_i$ denotes the frequency of the ancestral allele at locus $i$.

**Load after a bottleneck**

We performed simulations of a bottleneck in a single panmictic population of constant size $N$. We first let the population evolve for 1000 generations such that mutation load reaches equilibrium. The population size then changes instantaneously to $N_B$ and remains constant for $T$ generations. After that, population size instantaneously changes back to $N$ individuals. Let $M_0$ denote the expected number of deleterious mutations in a population that did not experience a bottleneck, and $M_B$ the number of deleterious mutations in a population experiencing a bottleneck. Figure S16 shows $(M_B - M_0)/M_B$, i.e., the relative excess (or deficiency) of deleterious mutations after the bottleneck. Negative values indicate that the total load decreased during the bottleneck.

## Human genomic data

### Excess of deleterious mutations in non-African populations

Human exomic diversity was inferred from the whole genome of 54 unrelated individuals from 11 populations sequenced by Complete Genomics at a depth of 51-89X coverage per genome [30]. We mapped SNPs to 504,378 coding exons of 19,086 autosomal human genes [Ensembl version 64, September 2011, [56]]. The derived and ancestral states of the SNPs were inferred from the comparison with the chimpanzee and orang-utan genomes, using the syntenic net alignments between hg19 and panTro2, and between hg19 and ponAbe2, both available on the UCSC platform [57]. We then kept the SNPs found to be polymorphic in 42 individuals from 3 African populations (4 Luhya from Webuye, Kenya; 4 Maasai from Kinyawa, Kenya; 9 Yoruba from Ibadan, Nigeria) and from 5 non-African populations (9 Utah residents with Northern and Western European ancestry from the CEPH collection; 4 Han Chinese from Beijing; 4 Gujarati Indians from Houston, Texas, USA; 4 Japanese from Tokyo; 4 Toscans from Italy). We predicted the functional consequences of SNP mutations using PolyPhen-2 [58] and we classified them as being either synonymous, non-synonymous neutral or non-synonymous deleterious.

### Diversity at non-neutral sites decreases with distance from Africa

Diversity at neutral and non-neutral sites was estimated by examining the HGDP-CEPH Human Genome Diversity Panel, which includes 660,918 SNPs typed in 53 populations from Africa, Middle-East, Europe, Central Asia, East Asia, Oceania and America [34]. We used GERP scores [59] to assess and quantify the level of evolutionary constraints acting on these SNPs. GERP scores can be defined as the number of substitutions expected under neutrality minus the number of substitutions observed at that position [36]. Positive scores larger than 2 represent a substitution deficit, which is expected for sites under selective constraints; while smaller scores, including negative values, indicate that a site is probably evolving neutrally [59]. GERP scores computed from the alignment of 35 mammals to the human genome reference sequence hg19 were downloaded from the UCSC platform [57,60] and could be assessed for 656,257 SNPs.

For each SNP locus and each of the population sample, we computed the expected heterozygosity as $H = 1 - \sum_i p_i^2$. In order to plot the average expected heterozygosity against the geographic distance from Addis Ababa, Ethiopia, a hypothetical but plausible point of origin for human expansion [16], we computed least-cost distances between the sampling location of HGDP populations and Addis-Ababa using PATHMATRIX [61].

## Analytical approximations under a simpler one-dimensional expansion model

Even though our model is relatively simple, it does not lead to the derivation of an analytical expression for the probability that a mutation becomes fixed at the edge of the expansion. However, we find simple approximations under a model where we separate migration from population growth such that migration only occurs once carrying capacity is reached. If $r \gg m$, the profile of the wave front will be very steep and individuals that colonize new territories will be only recruited from a few demes at the edge of the range. We therefore model the expansion process in the following way. First, neighboring demes exchange migrants such that $Km/2$ individuals move from deme $d_f(t)$ to deme $d_f(t)+1$, where $d_f(t)$ denotes the deme at the wave front in generation $t$. Then all demes reproduce and the population in the newly colonized deme grows exponentially until it reaches carrying capacity. We assume that selection acts through differential viability only during this growth phase.

In this model, the deme at the edge of the range evolves independently from demes in the range core, since migration rate *m* is used here only to define the number of founders in the newly colonized deme. We can thus study the dynamics of the wave front by simply tracking the dynamics of allele frequencies in deme $d_f$. The resulting model is equivalent to a model of repeated bottlenecks in a single population [62], and similar to a recent modeling of a range expansion as a series of founder effects [12].

**Probability of fixation**

We use a diffusion approximation to calculate the probability of fixation of a new mutation [24]. Note that fixation in our model means that the mutation becomes fixed at the front of the expansion wave, i.e., we do not consider the fate of mutations that fall back into the wake of the wave.

First, we consider a mutation that is introduced when demes are at carrying capacity and before migrants colonize a new deme. We then generalize our results to mutations that occur at an arbitrary stage of the colonization process. The selective advantage of an individual carrying a single copy of the mutation is denoted $s$. Let $\Delta x$ denote the change in mutant allele frequency $x$ during the colonization of a new deme and let $E[\Delta x]$ and $V[\Delta x]$ denote its mean and variance, respectively. The probability of fixation of the mutant allele is then given by

$$p = \frac{\int_0^{x_0} g(x)\,dx}{\int_0^1 g(x)\,dx},$$

where $x_0$ is the initial frequency of the mutation and $g(x) = \exp\left(-\int \frac{2E[\Delta x]}{V[\Delta x]} dx\right)$.

During exponential growth, selection deterministically changes the mutant frequency to

$$x' = \frac{x \exp[(r+s)T]}{x \exp[(r+s)T] + (1-x)\exp(rT)},$$

where $T \approx \log(2\ m^{-1})/r$ is the length of the growth phase, $r$ denotes the growth rate of the population, and $m$ is the migration rate (i.e., $m/2$ is the fraction of a deme`s population that colonizes a new deme). Binomial sampling of individuals during migration does not change the expected frequency of the mutation but it increases sampling variance around *x'*. Hence, the expected change in allele frequency is simply

$$E[\Delta x] = x - x'$$

and the variance of $\Delta x$ is

$$V[\Delta x] = \frac{1}{2K\frac{m}{2}} x'(1-x'),$$

where *Km* is the number of genes founding the new deme on the wave front. Under weak selection (i.e, $|s| \ll 1$), we find that the fixation probability is

$$p \approx \frac{\exp(4s_e K x_0) - 1}{\exp(4s_e K) - 1},$$

where $s_e = s\frac{m}{2}\frac{\log(2/m)}{r}$ can be considered as an effective selection coefficient for a mutation occurring on the wave front. If the mutation appears when the deme is at carrying capacity, $x_0 = \frac{1}{2K}$ and we find that the fixation probability is

$$p \approx \frac{\exp(4s_e) - 1}{\exp(4s_e K) - 1}. \qquad [2]$$

Note that Eq. [2] is equivalent to the probability of fixation of a mutation with effect $s_e$ in a single panmictic population of constant size *K* [24]. Hence, if $|Ks_e| \ll 1$, random genetic drift will dominate selection at the range margin. Comparison of Eq. [2] with individual based simulations shows that our approximation is very accurate for a broad range of parameters (Figure S7).

We next consider a mutation with effect $s$ that occurs $t$ generations after the initial colonization of a new deme and denote the probability of fixation of this mutation by $p(s,t)$. At the edge of the range, the number of individuals in a deme $t$ generations after it has been colonized is $N(t) = (m/2)K\exp(rt)$. If the mutation appears in generation $t < T$, its expected frequency in generation $T$ (i.e., when the deme at the edge is at carrying capacity, and before colonization of a new deme) will be

$$x_0(t) = \frac{\frac{1}{2N(t)}\exp[(r+s)(T-t)]}{\frac{1}{2N(t)}\exp[(r+s)(T-t)]+(1-\frac{1}{2N(t)})\exp[r(T-t)]},$$

and using Eq. [2] we immediately get the probability of fixation

$$p(s,t) \approx \frac{\exp(2s_e 2Kx_0(t))-1}{\exp(4s_e K)-1}.$$

**Change in mean fitness at the wave front**

We next calculate an approximation for the expected relative change in mean fitness per generation. We assume that the fitness on the wave front changes upon the successive fixation of new mutations, and that new mutations are either lost or fixed at any locus before the next mutation arises. In addition, we assume that we can treat loci independently (e.g., that recombination between loci is high).

We first consider the establishment of new mutations at a single locus. Let $u$ denote the genome wide mutation rate and $n$ the number of loci. Because mutations occur uniformly across the genome, the mutation rate per locus is $u/n$. If this mutation rate is sufficiently small, the probability that a new mutation occurs in generation $t$ is approximately $2N(t)un$. Let $\varphi(s)$ denote the probability that a mutation has effect $s$. Then the joint probability that a mutation with effect $s$ occurs in generation $t$ and then goes to fixation is

$$P(s,t) = 2N(t)\frac{u}{n}\varphi(s)p(s,t),$$

where $p(s,t)$ is again the probability of fixation of a mutation that occurs in generation $t$. A mutation that becomes fixed changes mean fitness by a factor $(1+s)(1+s) = 1+2s+s^2$. Because we consider only contributions to mean fitness by mutations that become fixed (i.e., we ignore the contributions of transient polymorphism), the expected effect on fitness of a mutation that occurs in generation $t$ is

$$\mu_s(t) = \sum_{\sigma}(2\sigma + \sigma^2)P(\sigma,t),$$

where $\sigma$ runs over all mutation effect sizes. If the distribution of fitness effects is continuous rather than discrete, the sum needs to be replaced by an integral, and $\varphi(s)$ then denotes the probability density of mutations with effect $s$. We then calculate the average value of $\mu_s(t)$ over the time *T* it takes for a deme to reach carrying capacity (and during which selection acts) as

$$\bar{\mu}_s = \frac{1}{T}\int_0^T \mu_s(\tau)d\tau.$$

We measure the change in mean fitness by the logarithm of the relative change in mean fitness in consecutive generations:

$$\omega = \log\left(\frac{\bar{w}_f(t+1)}{\bar{w}_f(t)}\right),$$

where the subscript *f* indicates that mean fitness is measured at the wave front.

Because fitness is multiplicative and loci evolve independently, the expected relative change in mean fitness per generation for $n$ loci is simply $(1+\bar{\mu}_s)^n$ and

$$\omega = \log\left[(1+\bar{\mu}_s)^n\right] = n\log(1+\bar{\mu}_s) \approx n\bar{\mu}_s. \qquad [3]$$

Eq. [3] is a key quantity to study the dynamics of expansion load: if $\omega < 0$, load will increase over time at a rate proportional to the absolute value of $\omega$. Furthermore, the expected mean fitness at the wave front evolves according to

$$E[\bar{w}_f(t+1) | \bar{w}_f(t)] = \bar{w}_f(t)(1+\bar{\mu}_s)^n. \qquad [4]$$

Here, $E[\bar{w}_f(t+1) | \bar{w}_f(t)]$ denotes the expected mean fitness at generation $t+1$ conditioned on the mean fitness at generation $t$.

To proceed further, we need to make assumptions about the distribution of fitness effects (DFE). Unless otherwise stated (see Figure S1), we henceforth assume that mutations have effects $\pm s$ and denote the fraction of mutations that are deleterious (i.e., the ones with effect $-s$) by $\varphi_d$.

Numerical investigations reveal that $\bar{\mu}_s$ is well approximated by $\mu_s(T)$ if selection is weak. Setting $\bar{\mu}_s = \mu_s(T)$ in Eq. [3], one can show that mean fitness on the wave front will decrease if $\varphi_d$ is larger than a critical threshold value

$$\varphi_{d,c} = \frac{1}{1+\exp(-4Ks_e)} \approx \frac{1}{2} + Ks_e, \qquad [5]$$

where $s_e$ is again the effective selection coefficient in our model. Figure S15 shows $\varphi_{d,c}$ as a function of $2Ks_e$.

## Acknowledgements

We thank Daniel Wegmann, Montgomery Slatkin, and Sarah Otto for their comments and useful suggestions. SP was supported by a US NSF grant DEB-0819901 to MK and by a Swiss NSF grant 31003A-143393 to LE.

# Figures

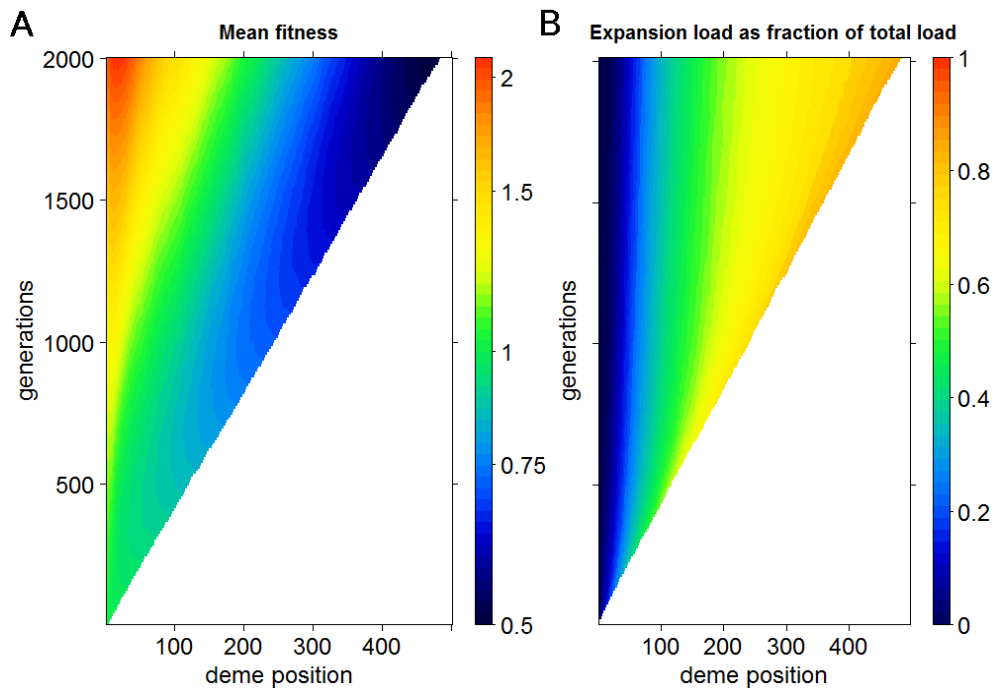

**Figure 1**. Evolution of population mean fitness (A) and expansion load (B) during a range expansion. A: Mean population fitness, normalized to 1 at the onset of the expansion. B: Fraction of the total load due to mutations originating either directly on the wave front or one deme away from it. Plots are average over 50 simulations. Results are for diploid individuals with *n* = 20 freely recombining regions. The effects of mutations are $s = \pm 0.005$, demes have a carrying capacity of $K$ = 100 diploid individuals, populations grow logistically at rate *r* = log(2), individuals migrate at rate *m* = 0.05 between adjacent demes, the genome wide mutation rate is *u* = 0.05, and the fraction of mutations that are deleterious is $\varphi_d$ = 0.9.

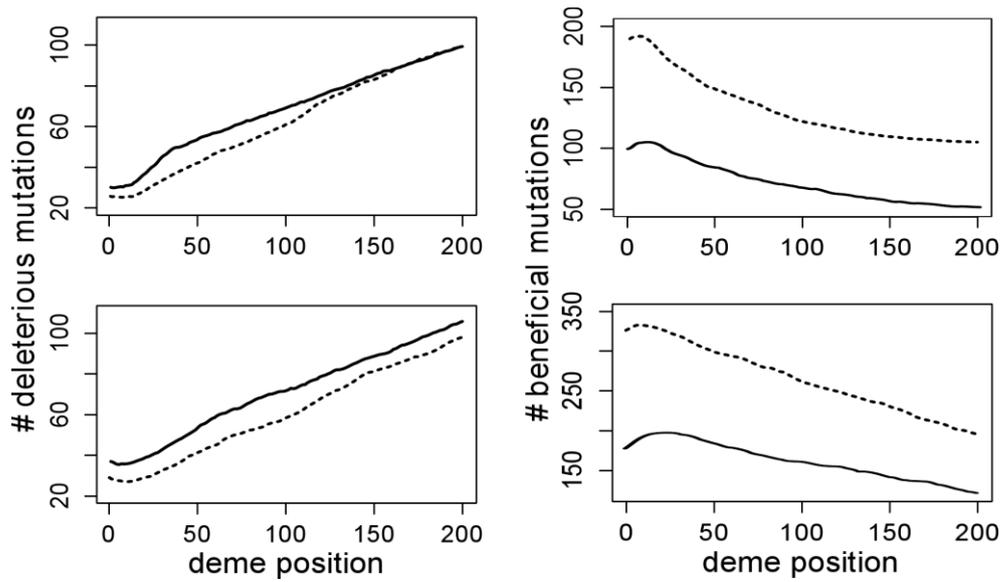

**Figure 2.** Spatial distribution of deleterious (left) and beneficial (right) mutations during linear range expansions. Solid lines show the average number of deleterious and beneficial mutations per individual over the first 200 demes of a 1D species range (cf. Figure 1) that has been colonized from the 5 leftmost deme either 1000 generations (top) or 2000 generations (bottom) after the onset of the expansion. Dashed black lines show the mean number of mutations during 2D linear expansions in a 10x200 deme habitat. Parameter values are as in Figure 1. Note the different scales in the top right and lower right panel.

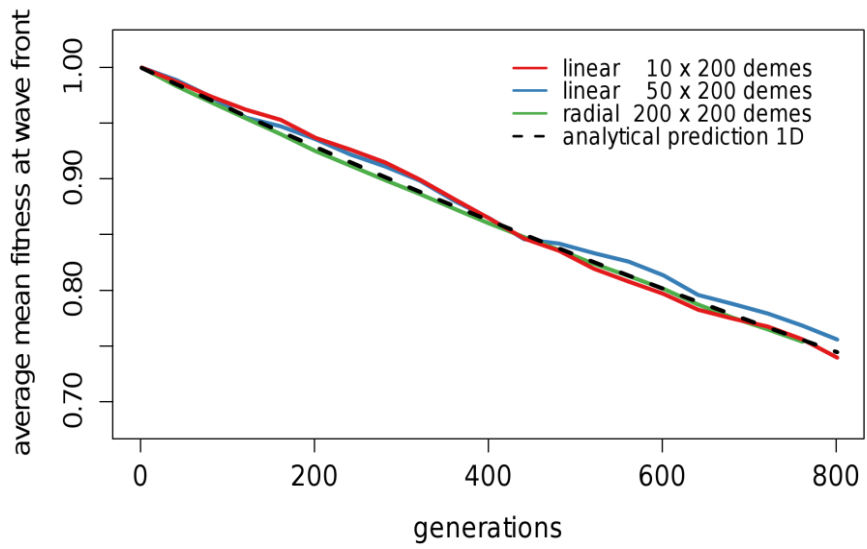

**Figure 3.** Decline of mean fitness at the wave front for different expansion types. The figure shows the average mean fitness on the whole wave front (calculated from 20 simulation runs). Parameter values are $K$ = 100, $r$ = log(2), $u$ = 0.05, $\varphi_d$ = 0.9, and $n$ = 20.

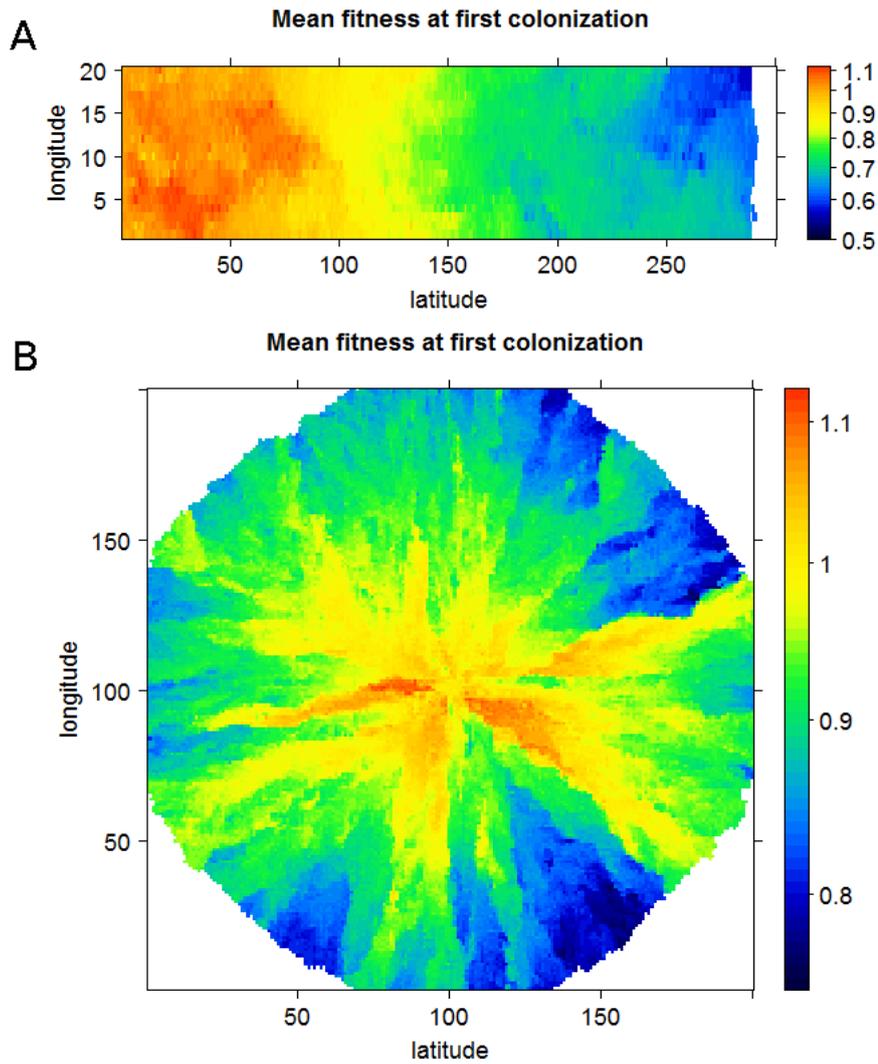

**Figure 4.** Evolution of the mean fitness at the wave front during an expansion in a 2D habitat. Mean fitness is shown for the generation at which each deme is colonized. A: Linear expansion starting from the 5 leftmost columns of demes, assumed to be at carrying capacity. B: Radial expansion starting from 25 demes at carrying capacity arranged in a 5x5 square in the middle of the habitat. In both cases, expansion happens after a burn in phase of 1000 generations; parameter values are as in Figure 1 ($K = 100$, $r = \log(2)$, $u = 0.05$, $\varphi_d = 0.9$, and $n = 20$).

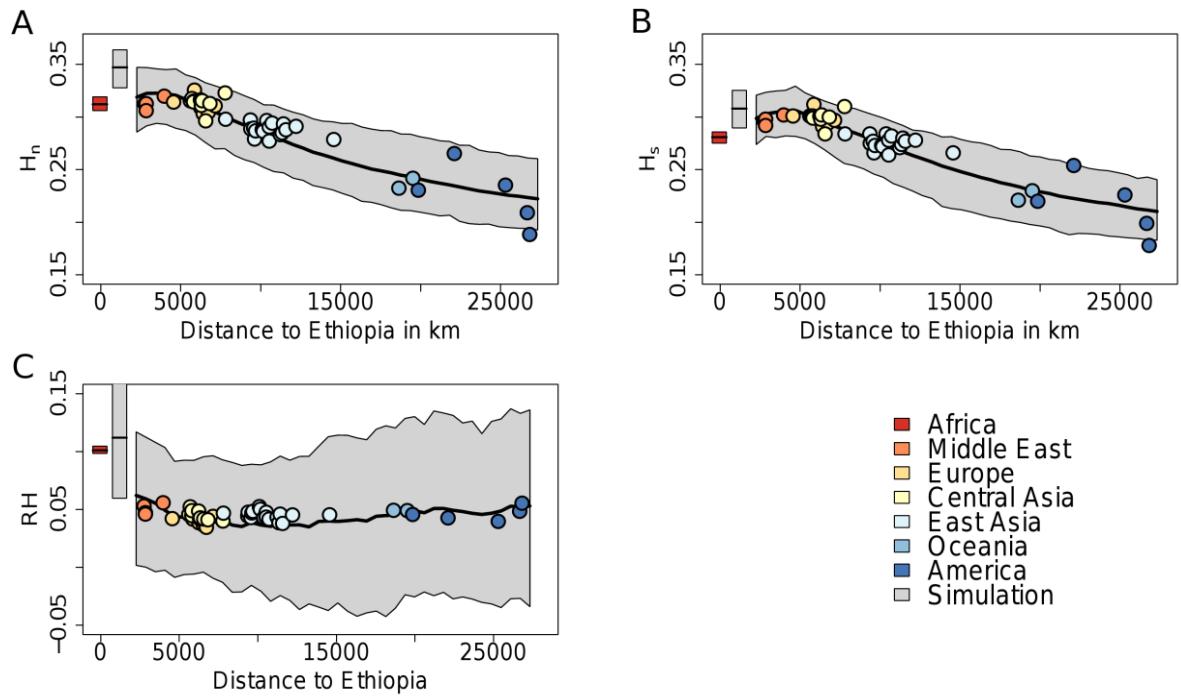

**Figure 5.** Comparison of HGDP SNP data with simulation results. Colored circles show average expected heterozygosity for human populations. Panel A shows the average heterozygosity of neutral sites and panel B the heterozygosity of sites under puryfing selection. Panel C shows the relative loss of heterozygosity *RH*. The red and gray boxplots on the left of each panel show the distribution of observed and simulated expected heterozygosities, in the African and ancestral populations, respectively. Solid black line shows the average over 100 simulation runs and the gray-shaded area indicates empirical 90% confidence intervals. Parameter values used in the simulations are *m* = 0.1, s = 0.002, *K* = 100, *r* = log(2), which have been chosen to produce a good fit with the observations, without real attempt at maximizing this fit.

# Tables

**Table 1**: Proportion of exomic sites with deleterious mutations in humans.

| Mutation frequencies | All sites Africans | Sign. | All sites non-Africans | Sign. | Private sites Africans | Sign. | Private sites non-Africans | Sign. | Shared sites | Sign. | All sites Afr vs. Non Afr Sign. | Private vs. all sites Afr. Sign. | Private vs. all sites Non Afr Sign. |
|---|---|---|---|---|---|---|---|---|---|---|---|---|---|
| All | 0.166 | *** [a] | 0.185 | *** [b] | 0.210 | *** [a] | 0.274 | *** [b] | 0.104 | *** [a] | *** | *** [c] | *** [d] |
| Rare (<=1/34) | 0.231 | ** [a] | 0.282 | *** [b] | 0.238 | *** [a] | 0.298 | *** [b] | 0.171 | *** [a] | *** | *** [c] | NS |
| Common (>1/34) | 0.131 | *** [b] | 0.127 | NS | 0.177 | *** [a] | 0.218 | * [b] | 0.093 | *** [a] | *** | *** [c] | ** [d] |

Significance levels are obtained by 1000 permutations of individuals between African and non-African groups: ***, $p<0.001$; **, $p<0.01$; *, $p<0.05$.
[a] Significantly smaller proportion of deleterious mutations than in a randomly mixed group of Africans and non-Africans. [b] Significantly larger proportion of deleterious mutations than in a randomly mixed group of Africans and non-Africans. [c] Smaller excess of deleterious mutations at private sites than expected by chance in a randomly mixed group of Africans and non-Africans. [d] Larger excess of deleterious mutations at private sites than expected by chance.

# Supporting Figures

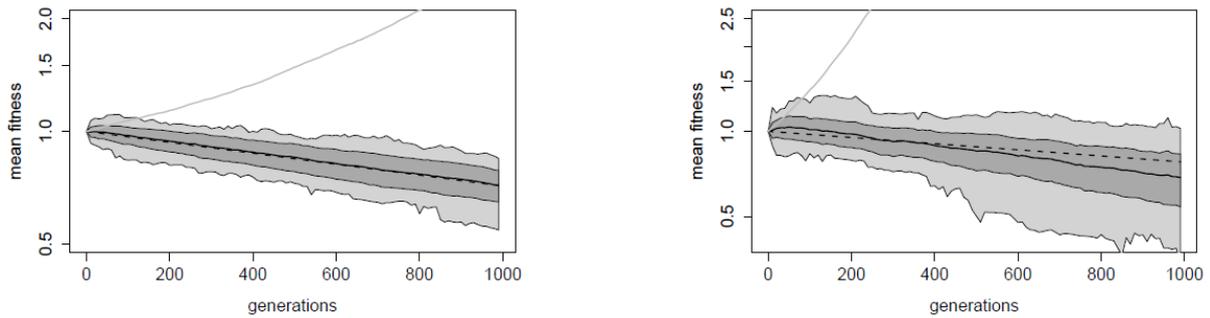

**Figure S1.** Changes in mean fitness on the wave front of an expanding population for different distribution of fitness effects. In each graph, the solid black line shows the simulated mean fitness at the wave front, dark gray area indicates two standard deviations of mean fitness, and light gray area the total range of observed simulated values. The solid gray line shows the average mean fitness in the core of the population, also obtained by simulations. Averages are taken over 50 simulations. Left: with probability 0.9 a mutation's effect is drawn from an exponential distribution with mean $s=-0.005$, and with probability 0.1 it is drawn from an exponential distribution with mean $s=0.005$. Right: discrete DFE where mutations with effect $s=-0.05$, $-0.005$, 0.005, and 0.05 occur with probability $\varphi(s)=0.09$, 0.81, 0.081, and 0.009, respectively. Results are for individuals with $n=20$ freely recombining regions. Other parameter values are $K=100$, $r=\log(2)$, $m=0.05$, and $u=0.05$.

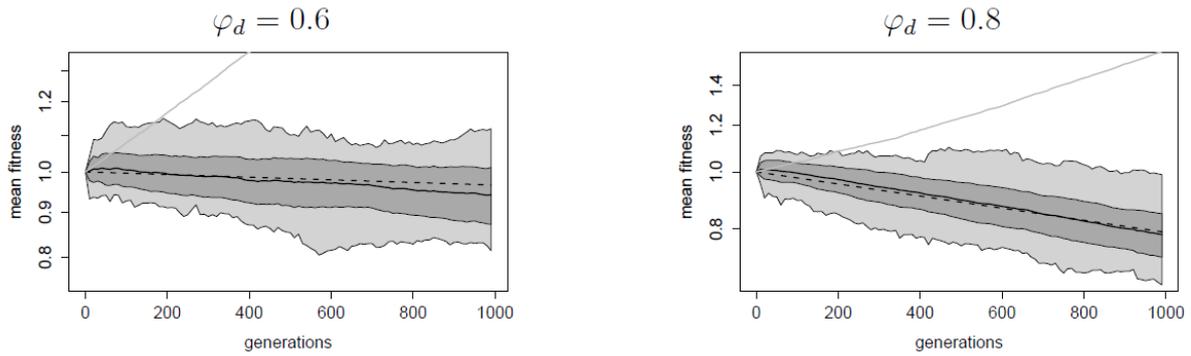

**Figure S2.** Changes in mean fitness on the wave front of an expanding population for different ratios of deleterious and beneficial mutations. The color code is as in Figure S1. Results are for individuals with $n=20$ freely recombining regions. Other parameter values are $s=\pm 0.005$, $K=100$, $r=\log(2)$, $m=0.05$, and $u=0.05$.

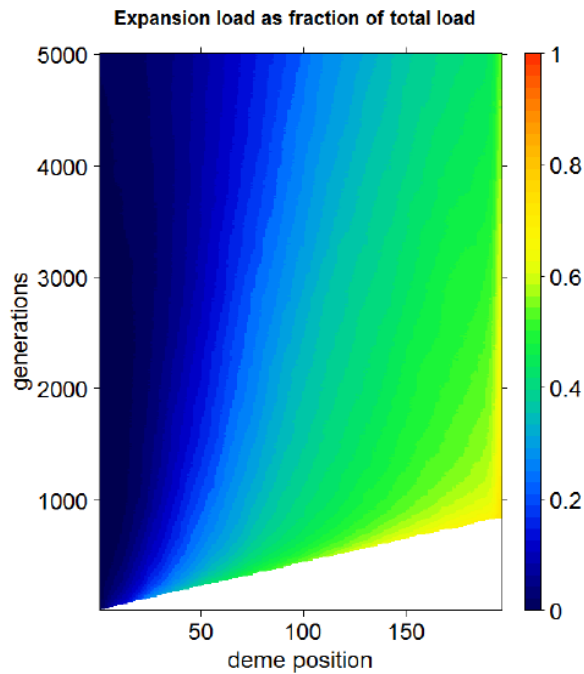

**Figure S3.** Fraction of total load that originated from the wave front (expansion load) for a finite range. Parameter values are as in Figure 1. Average is taken over 50 simulations.

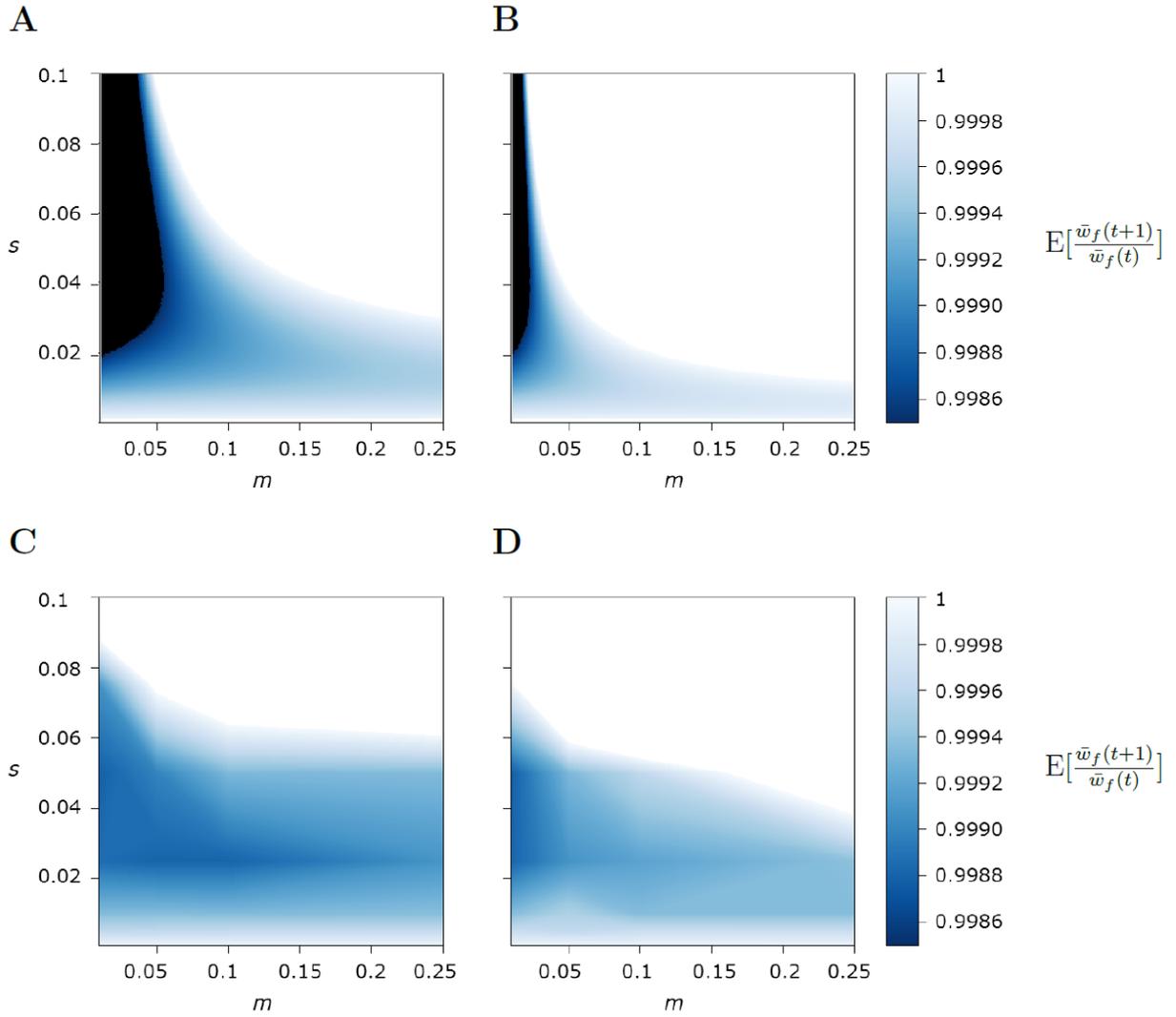

**Figure S4.** Change in mean fitness on the wave front. The contour plot shows the expected change in mean fitness at the wave front, $E[\bar{w}_f(t+1)/\bar{w}_f(t)]$, as a function of $m$ and $s$ for $K=100$ (A and C) and $K=250$ (B and D). Mutations have effect $\pm s$. Panels A and B are obtained from the analytical approximation [4], and panels C and D are obtained from simulation of the more complex expansion model. In the white area, $E[\bar{w}_f(t+1)/\bar{w}_f(t)] > 1$, and the black area indicates values smaller than the minimum of the color key. Remaining parameter values are $\varphi_d = 0.9, r = \log(2),$ and $u = 0.05$.

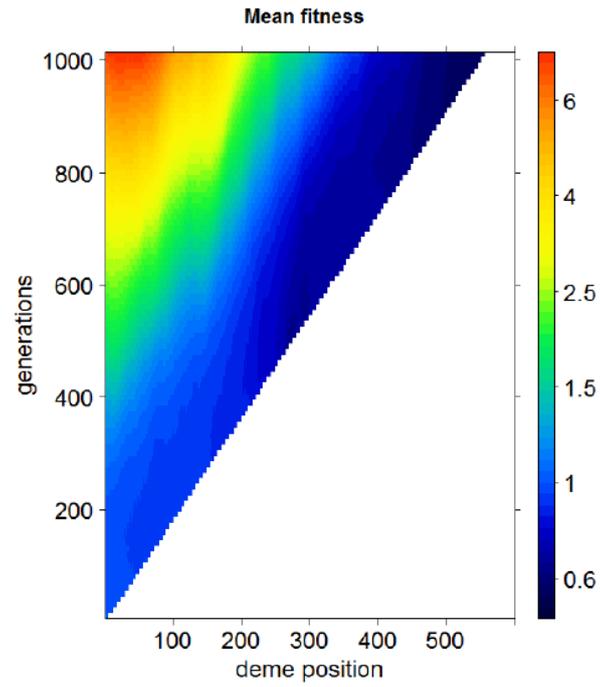

**Figure S5.** Evolution of population mean fitness with large local carrying capacities. Results are for individuals with $n = 20$ freely recombining regions. Parameter values are $s = \pm 0.01$, $K = 1000$, $r = \log(2)$, $m = 0.25$, $u = 0.05$, and $\varphi_d = 0.9$.

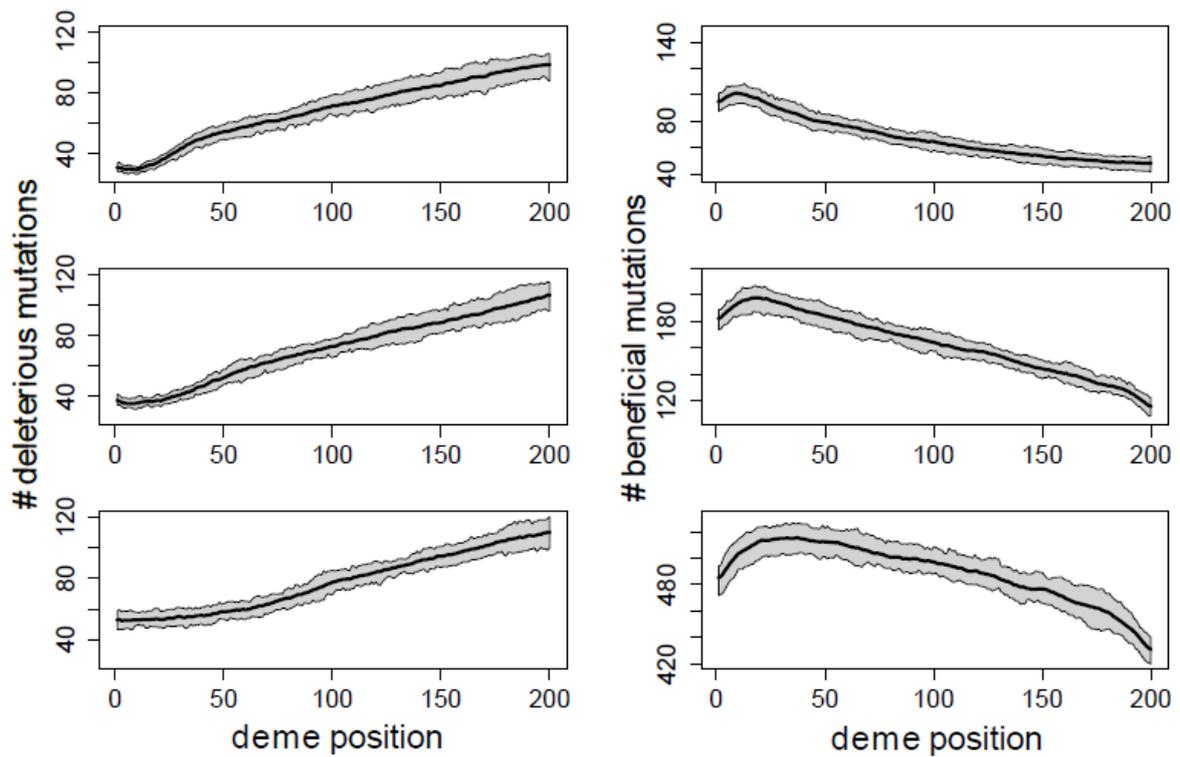

**Figure S6.** Spatial distribution of mutations under after a range expansion in a linear habitat restricted to 200 demes. The left and right panels show the number of deleterious and beneficial mutations in the 200 demes of the species range, respectively. Top: 1000 generations after the onset of the expansion, Middle: 2000 generations after the onset of the expansion. Bottom: 5000 generations after the onset of the expansion. Solid black line shows mean number of mutations and grey areas indicate the lower and upper quartiles. Average is taken over 50 simulations. Parameter values are as in Figure 1.

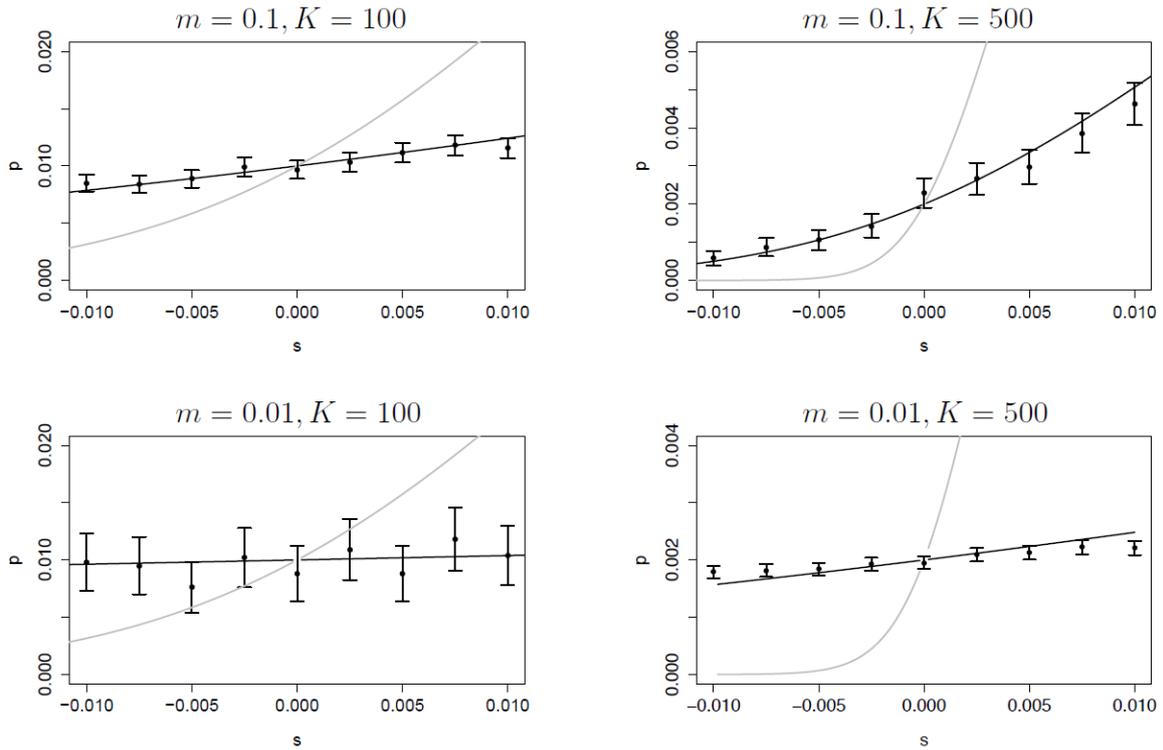

**Figure S7.** Probability of fixation of mutations at the wave front. Solid black line shows the analytical prediction [2], black dots show results from simulations of a more complicated expansion model (see *Material and methods*), and whiskers represent empirical 99% confidence intervals obtained by simulations. For comparison, the solid gray line shows the probability of fixation of a mutation in a single panmictic population. In the simulations the five leftmost demes were at carrying capacity and all other demes were empty before the onset of the expansion. A single copy of the mutation was present in the deme at the edge of the expansion, i.e., $x_0 = 1/2K$. In all cases $r = \log(2)$.

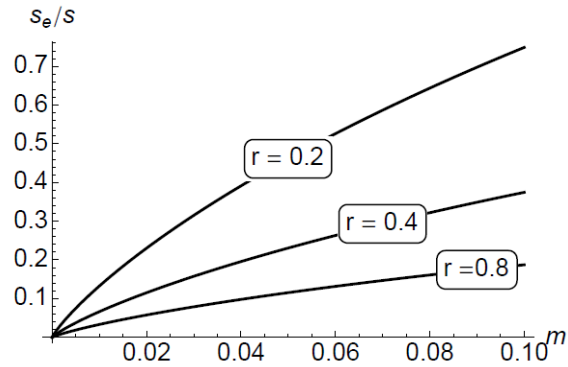

**Figure S8.** Ratio of the effective selection coefficent at the wave front, $s_e$, and the actual selection coefficient, $s$.

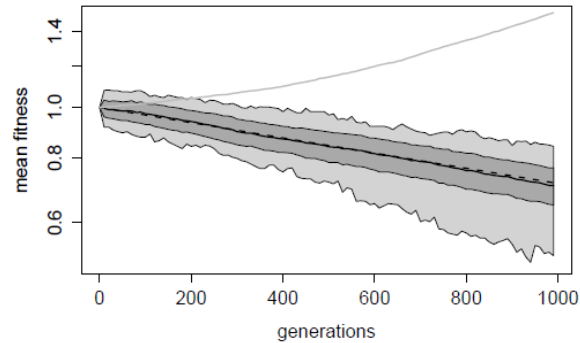

**Figure S9.** Changes in mean fitness on the wave front of an expanding population with long distance dispersal. Dispersal distances are drawn from a shifted exponential distribution such that the mean dispersal distance is 5 demes. The color code is as in Figure S1. Results are for individuals with n = 20 freely recombining regions. Other parameters are $s = \pm 0.005,$, $K = 100$, $r = \log(2)$, $m = 0.05$, $\varphi_d = 0.9$ and $u = 0.05$.

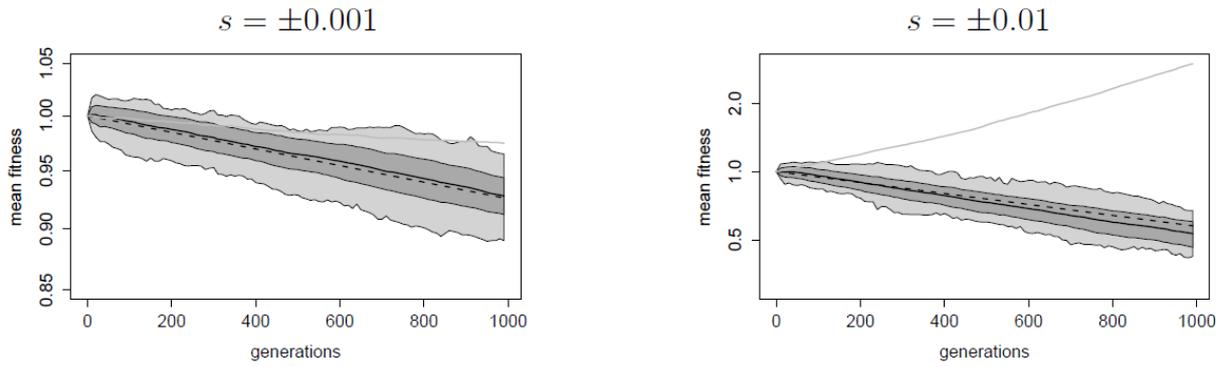

**Figure S10.** Changes in mean fitness on the wave front of an expanding population. The color code is as in Figure S1. In each graph, the solid black line shows the simulated mean fitness at the wave front, dark gray area indicates two standard deviations of mean fitness, and light gray area the total range of observed simulated values. The solid gray line shows the average mean fitness in the core of the population, also obtained by simulations. Averages are taken over 50 simulations. The dashed line shows the analytical approximation [4]. Results are for individuals with $n = 20$ freely recombining regions. Other parameters are $m = 0.05, K = 100, r = \log(2), \varphi_d = 0.9,$ and $u = 0.05$.

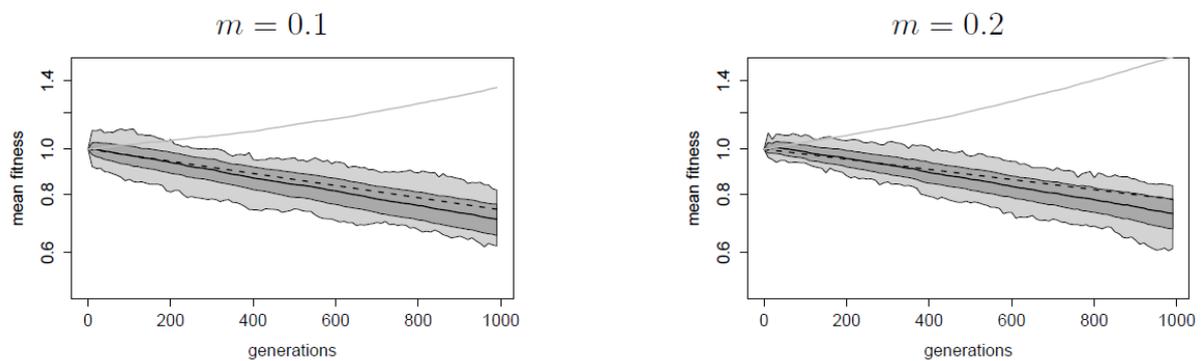

**Figure S11.** Changes in mean fitness on the wave front of an expanding population for different migration rates. The color code is as in Figure S1. Results are for individuals with $n = 20$ freely recombining regions. Other parameters are $s = \pm 0.005, K = 100, r = \log(2), \varphi_d = 0.9,$ and $u = 0.05$.

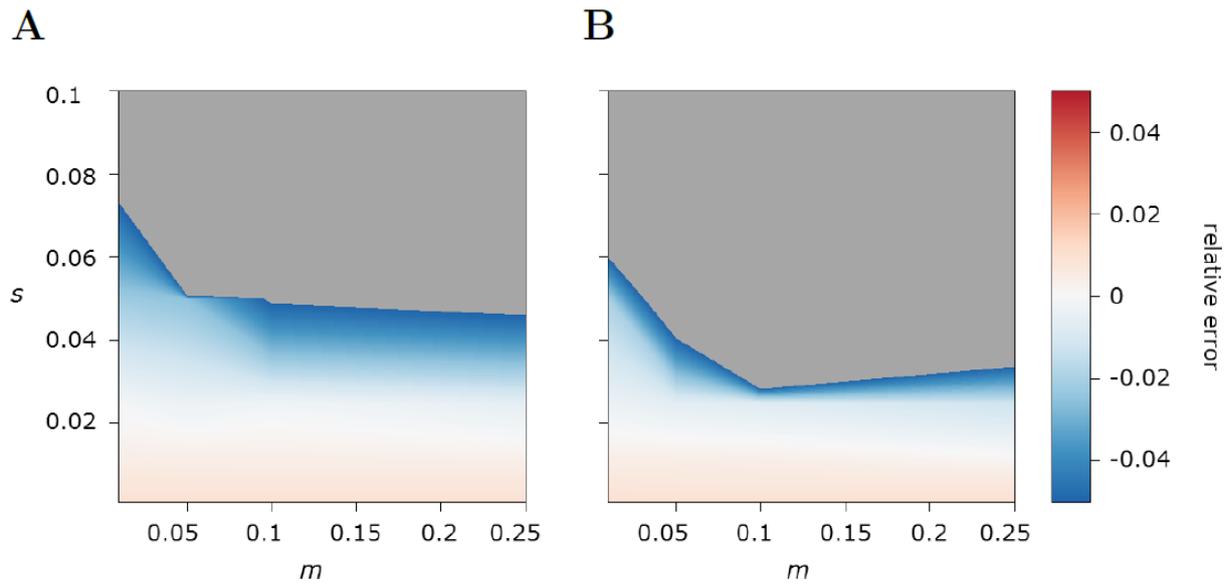

**Figure S12.** Relative error of the analytical approximation [4] as a function of $m$ and $s$ for $K=100$ (A) and $K=250$ (B). Positive values (red) mean that the analytical approximation overestimates the decline of mean fitness at the wave front, negative values (blue) mean that the analytical approximation underestimates the decline of mean fitness at the wave front. In the gray-shaded area the relative error is larger than 5%. Parameter values are as in Figure S4.

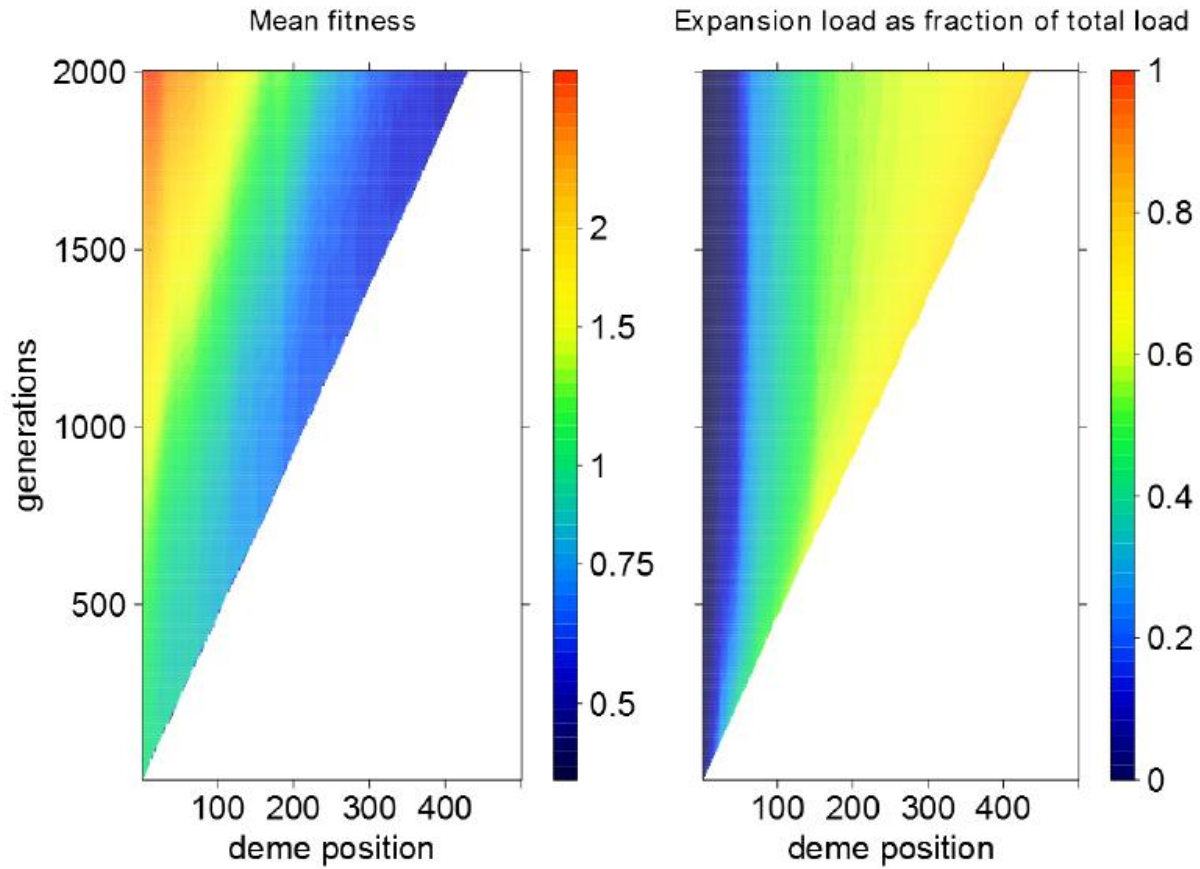

**Figure S13.** Evolution of population mean fitness and expansion load during a range expansion in a two-dimensional habitat. A: mean population fitness (normalized to 1 at the onset of the expansion). B: fraction of the total load due to mutations originating on the wave front. A mutation is considered to have originated at the front if it first appeared in an individual living in the deme currently at the edge of the expansion or one deme behind it. The habitat is 10 demes wide. The plots show averages over the width of the habitat. Results are for individuals with n = 20 freely recombining regions. Parameter values are $s = \pm 0.005$, $K = 100$, $r = \log(2)$, $m = 0.05$, $u = 0.05$, $\varphi_d = 0.9$.

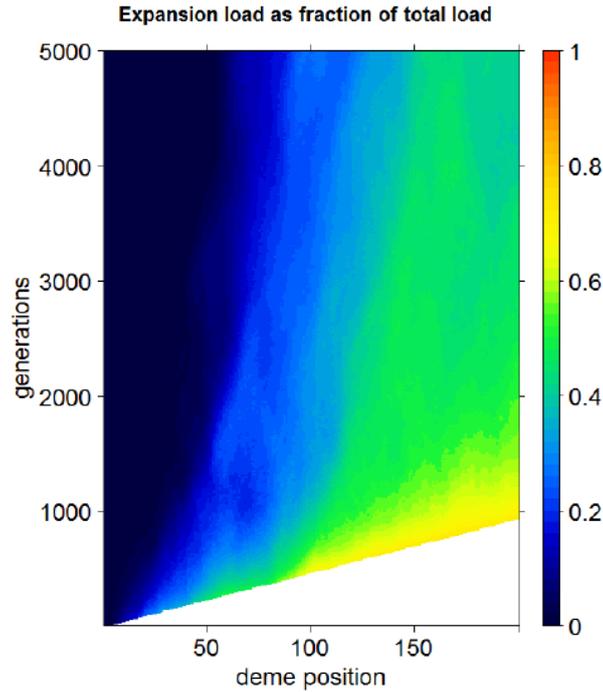

**Figure S14.** Fraction of total load that originated from the wave front (expansion load) for a linear range expansion in a finite two-dimensional habitat. The habitat is 20x200 demes. The plot shows the average over the width of the habitat. Results are for individuals with n = 20 freely recombining regions. Parameter values are $s = \pm 0.005$, $K = 100$, $r = \log(2)$, $m = 0.05$, $u = 0.05$, $\varphi_d = 0.9$.

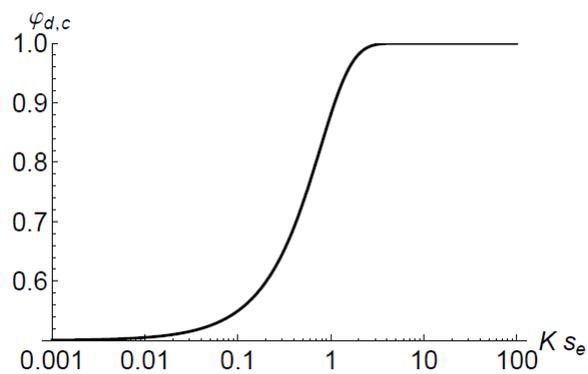

**Figure S15.** Critical value of $\varphi_d$ as a function of $2Ks_e$. Mean fitness at the wave front decreases for values of $\varphi_d$ above the solid lines.

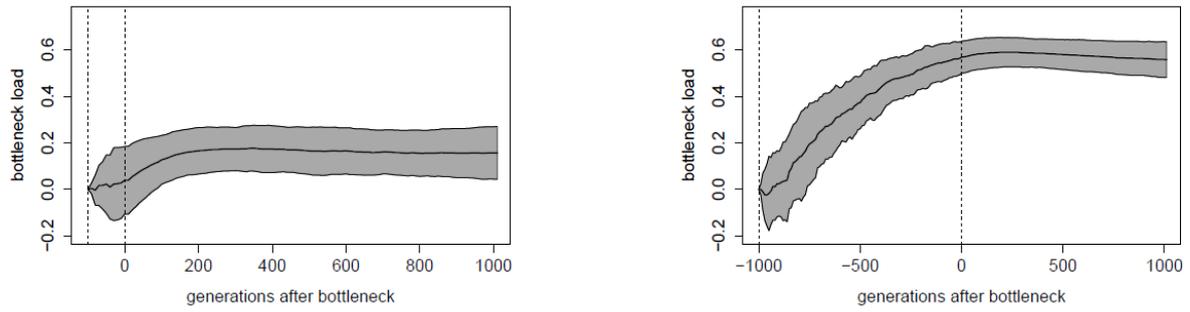

**Figure S16.** Fraction of the total load that is established during a single bottleneck. Thick black line shows the average over 50 realizations and gray area indicates two standard deviations. Vertical dashed lines indicate beginning and end of the bottleneck. Results are for $n=20$ freely recombining regions. Parameter values are $s=\pm 0.005$, $u=0.05$, $\varphi_d=0.9$, $N=10000$, $N_B=100$, $T=100$ (left), and $T=1000$ (right).

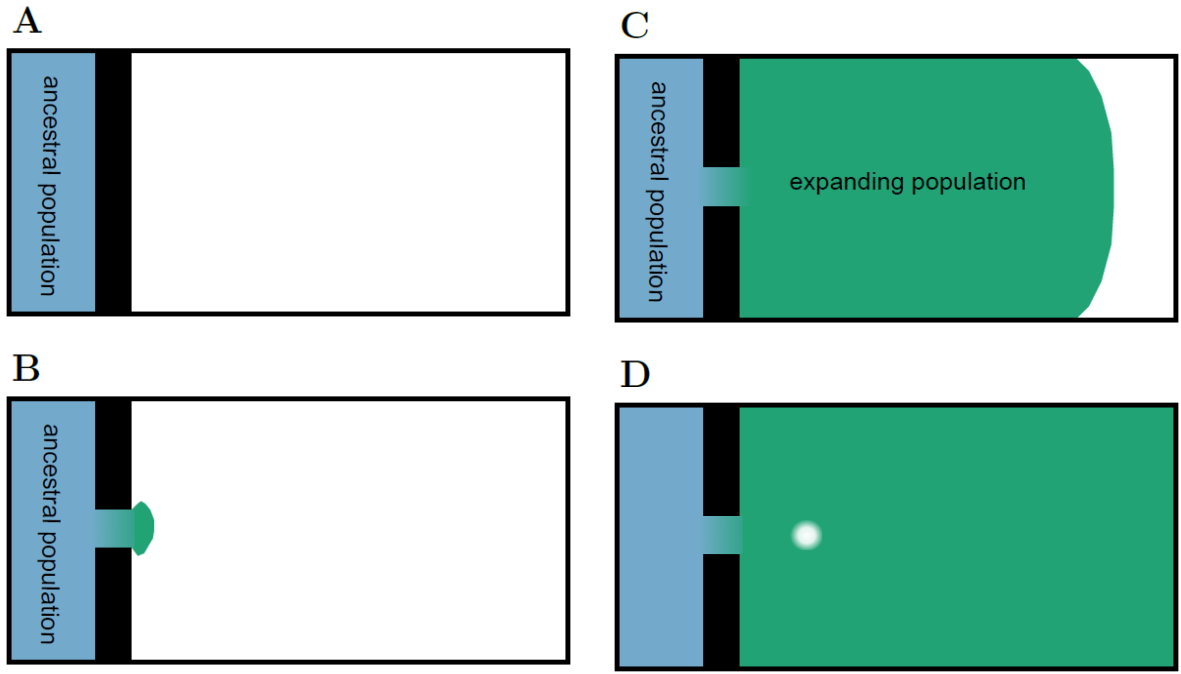

**Figure S17.** Sketch of the expansion model with a spatial bottleneck we used to simulate the evolution of heterozygosity during a linear 2D expansion. Panel A shows the ancestral population (blue) separated from the empty habitat by a migration barrier (black). Panel B shows the onset of the expansion and panel C the colonization of the empty habitat by the expanding population (green). Panel D shows the whole metapopulation after the colonization is complete. The white dot indicates the population from which individuals were sampled to determine the loci used in the correction for ascertainment bias.

## Supporting Video Captions

**Supporting Video 1:** Evolution of average expected heterozygosity and mean fitness. The video shows an example of the simulations that were used to obtain Figure 5.